# Stress effect on magnetoimpedance (MI) in amorphous wires at GHz frequencies and application to stress-tunable microwave composite materials

L. V. Panina[a], S. I. Sandacci, and D. P. Makhnovskiy[b]

*School of Computing, Communications and Electronics, University of Plymouth,*

*Drake Circus, Plymouth, Devon PL4 8AA, United Kingdom.*

*Abstract* − The effect of tensile stress on magnetoimpedance (MI) in CoMnSiB amorphous wires at microwave frequencies (0.5–3 GHz) is investigated both experimentally and theoretically. In the presence of the dc bias magnetic field of the order of the anisotropy field, the impedance shows very large and sensitive change when the wire is subjected to a tensile stress: 100% and 60% per 180 MPa for frequencies 500 MHz and 2.5 GHz, respectively. It is demonstrated that this behavior owes mainly to the directional change in the equilibrium magnetization caused by the applied stress and field, which agrees well with the theoretical results for the surface impedance. This stress effect on MI is proposed to use for creating microwave stress-tunable composite materials containing short magnetic wires. The analysis of the dielectric response from such materials shows that depending on the stress level in the material, the dispersion of the effective permittivity can be of a resonant or relaxation type with a considerable change in its values (up to 100% at 600 MPa). This media can be used for structural stress monitoring by microwave contrast imaging.

[a] Electronic mail: lpanina@plymouth.ac.uk

[b] Electronic mail: dmakhnovskiy@plymouth.ac.uk

**I. Introduction**

Amorphous magnetic wires having Co-alloy compositions exhibit very large change in complex-valued impedance, when subjected to a high frequency current and dc magnetic field.[1-3] This effect known as magnetoimpedance (MI) is caused by the dependence of high frequency current distribution in the wire on its magnetic properties, in particular, equilibrium magnetic structure and dynamic transverse magnetization (with respect to the current flow). It appears that the strongest impact on the MI sensitivity is related with the directional change in the static magnetization. CoFeSiB amorphous wires have a circumferential (or helical) anisotropy due to coupling between a negative magnetostriction and residual stress. Applying the axial magnetic field rotates the equilibrium magnetization $\mathbf{M}_0$ towards the axis. Within the field range, where the reorientation in $\mathbf{M}_0$ takes place, which is defined by the anisotropy field $H_K$, the MI sensitivity (impedance change ratio per Oersted) runs into more than 100%/Oe at Megahertz frequencies[1-4] and more than 50%/Oe at Gigahertz frequencies.[5-7] In the latter case, the change in impedance is entirely due to the rotation of $\mathbf{M}_0$ by the external field $H_{ex}$ since the operational frequencies are much higher than the frequency of ferromagnetic resonance for $H_{ex} < H_K \approx 1-5$ Oe. Owing to magnetostrictive origin of the anisotropy, applying a stress or a torque also can cause a sensitive change in the wire magnetic structure and very large variations in impedance. It was reported that CoSiB amorphous wires with a helical anisotropy have a strain gauge factor of 1200–2000 at 1–10 MHz.[8] These unique magnetic and ac transport properties make amorphous wire elements suitable for a wide range of applications: magnetic sensors for low field detection,[9,10] highly sensitive stress sensors,[11,12] radio frequency wireless sensors[13] and tunable microwave composite materials.[14-16] The present paper further advances this field of research by investigating the effect of applied tensile stress on MI in CoMnSiB amorphous wires at microwave frequencies and proposing new stress-sensitive microwave materials. The relative change in impedance of about 60% for applied stress of 180 MPa at frequencies as high as 2–3 GHz has been demonstrated, which agrees well with the theoretical results. This effect could be of a considerable interest for applications in smart composite materials filled with short pieces of ferromagnetic wires for remote stress monitoring, where the effective permittivity depends on stress or strain in a narrow frequency band with a characteristic frequency of few GHz.

In general, artificial material with conducting fibers present a considerable interest since their dielectric response characterized by an effective permittivity, $\varepsilon_{eff}$ (which can be introduced, at least, in the far-field region) exhibits various dispersive behaviors.[15,17,18] The composite with short wires demonstrates a resonance type of $\varepsilon_{eff}$. The wire inclusions perform as "elementary scatterers" (microantennas), when the electromagnetic wave irradiates the composite and induces an electrical dipole moment in each inclusion. At the antenna resonance the dispersion of the effective permittivity is strongly modulated by the contribution from wire dipole moments even for very small concentrations of inclusions. In this frequency range, the variations in the wire impedance due to the MI effect may produce a considerable change in the current density distribution within the wires and, consequently, in their induced dipole moment. The resulting effect of the impedance change will be then a considerable change in $\varepsilon_{eff}$ in this frequency range.[15] Therefore, the proposed composite demonstrates both the tunable and resonance properties (selective absorption). In the case of the wire inclusions having stress-dependent MI, the effective permittivity becomes sensitive to internal strains in the host material caused by various factors such as external load, temperature gradient, material fatigue, etc. The result of modeling based on stress-impedance data shows the variation in the dielectric loss (imaginary part of $\varepsilon_{eff}$) of such composite more than twice under the effect of stress of 650 MPa. For practical applications, stress imaging can be achieved by microwave contrast scanning the proposed media and analyzing the reflected signals in adjacent areas.

**II. Analysis of stress effect on MI in amorphous wires**

An ac voltage response from a ferromagnetic specimen is convenient to represent in terms of the surface impedance tensor $\hat{\varsigma}$ that relates tangential electric **e** and magnetic **h** fields at the surface:[1]

$$\mathbf{e} = \hat{\varsigma}(\mathbf{h} \times \mathbf{n}), \tag{1}$$

where **n** is a unit vector along the inward normal to the surface. In general, Eq. (1) can be used as a boundary condition in determining the field outside the metal, which will be used later when solving a scattering problem at a ferromagnetic wire. For a wire geometry carrying a high frequency current $i = i_0 \exp(-j\omega t)$, the voltage $V$ measured across it is given by the longitudinal component $\varsigma_{zz} = e_z / h_\varphi$:

$$V = l\frac{2i}{ca}\varsigma_{zz}. \tag{2}$$

Here $e_z$ and $h_\varphi = 2i/ca$ are the axial electric and circular magnetic fields at the surface, $l$ is the wire length, $a$ is the wire radius, and $c$ is the velocity of light (Gaussean units are employed). Strictly speaking, Eq. (2) is valid for a wire with a uniform magnetization but can be applied for a multidomain wire provided that $\varsigma_{zz}$ is averaged over domains and the domain wall displacements are damped. In the case of a strong skin effect, the following equation is held:[1]

$$\varsigma_{zz} = \frac{c\rho(1-j)}{4\pi\delta_0}\left(\sqrt{\mu}\cos^2\theta + \sin^2\theta\right). \tag{3}$$

Here $\rho$ is resistivity, $\delta_0 = c(\rho/2\pi\omega)^{1/2}$ is a non-magnetic skin depth, $\mu$ is the circular permeability describing the magnetization precession around the equilibrium magnetization $\mathbf{M}_0$ having an angle $\theta$ with the wire axis (z-axis). Eq. (3) demonstrates that the longitudinal surface impedance depends on both the dynamic permeability $\mu$ and the static magnetization angle $\theta$. The parameter $\mu$ has a very broad dispersion region, as shown in Fig. 1, where the permeability plots obtained from the liberalized Landau-Lifshitz equation for a uniform magnetization precession are given. At high frequencies higher than the frequency of the ferromagnetic resonance, μ changes very little with both the external axial magnetic field $H_{ex}$ and anisotropy field $H_K$, although preserving its relatively high values ($|\mu| \approx 40$ at 2-3 GHz for Co-based wires with $H_K = 2.5$ Oe). In this case, the dependence of the impedance on magnetic properties is entirely determined by the static magnetization angle θ.

In amorphous ferromagnetic alloys, the preferable magnetization direction is set by the combined effect of the shape anisotropy and the magnetoelastic anisotropy arising from the coupling between magnetostriction λ and internal stresses σ. The sign and value of λ depend on the alloy composition. Thus, negative magnetostriction, which is proven to be essential for the MI effect, is typical of Co-based alloys having relatively small λ of about $-3 \cdot 10^{-6}$ that can be reduced down to $-10^{-7}$ by small additions of Fe or Mn. The internal stresses (axial, radial, and azimuth) are induced during the fabrication process and can be controlled by further post production treatment like drawing and annealing. The initial process of quenching produces a substantial distribution of σ over radius due to a strong thermal gradient as the solidification front proceeds inward to the center of the wire.[19] This results in a complicated core and sheet domain structure. In the case of glass-coated amorphous wires, the stresses are also produced due to the contraction of the two materials (metal and glass) having different thermal expansion coefficients,[20,21] which may further complicate the domain structure. However, the experimental hysteresis loops are nearly rectangular for $\lambda > 0$ and nearly linear for $\lambda < 0$, suggesting that the domain structure is rather uniform.[22] It appears that an additional large tensile axial stress is produced by the drawing of the wire, which gives a dominant contribution to the magnetostrictive energy.

Further we consider glass coated microwires and assume that in general the internal stress consists of an axial tension ($\sigma_1$) and a torsion which is a combination of a tension and compression with equal intensity $\sigma_2$ at $90^0$ to each other and at $45^0$ to the wire axis. Both of these stresses can be produced by drawing processes and typically $\sigma_1 >> \sigma_2$. This is certainly a simplified model; nevertheless, it describes quantitatively the experimental results on hysteresis loops and impedance. For $\lambda < 0$, the easy anisotropy is directed in a helical way having an angle $\alpha < 45^0$ with respect to the circumference. The static magnetization direction can be changed by the application of the external axial magnetic field $H_{ex}$ and stress. Here we consider the effect of tensile stress $\sigma_{ex}$. For such stress configuration, the magnetoelastic energy is written as $U_{me} = -(3/2)\lambda\left((\sigma_1 + \sigma_{ex})\cos^2\theta + \sigma_2 \cos^2(\theta - 45°) - \sigma_2 \cos^2(\theta + 45°)\right)$, which is convenient to represent in the form of an effective uniaxial anisotropy:

$$U_{me} = -\frac{3}{2}|\lambda|\tilde{\sigma}\sin^2(\alpha + \theta), \tag{4}$$

$$\alpha = \frac{1}{2}\tan^{-1}\left(\frac{2\sigma_2}{\sigma_1 + \sigma_{ex}}\right), \quad \tilde{\sigma} = \frac{\sigma_1 + \sigma_{ex}}{\cos(2\alpha)}.$$

The stable direction of $\mathbf{M}_0$ is found by minimizing the magnetostatic energy $U = U_{me} - M_0 H_{ex} \cos(\theta)$.

For a large stress effect on MI, the ability of $\sigma_{ex}$ to change the direction of $\mathbf{M}_0$ is needed. In the case of $H_{ex} = 0$, this is possible if the anisotropy axis has a helical angle nearly equal to $45^0$, which is achieved by a proper annealing (current annealing or annealing under torsion) to establish a substantial frozen-in torsion ($\sigma_2 >> \sigma_1$, $\alpha \approx 45^0$ at $\sigma_{ex} = 0$). Another possibility of the directional change in the magnetization under the external stress is to use the axial field bias. In the presence of $H_{ex}$ the magnetization rotates towards the wire axis whereas the application of $\sigma_{ex}$ strengthens the circumferential anisotropy (for $\lambda < 0$) and hence acts in the opposite way. Then, the magnetization rotates back to the circular direction. In this paper, we utilize the assistance of $H_{ex}$ to realize very large impedance change when the external tensile stress is applied.

### III. Experimental

The experimental results are provided for $Co_{68.5}Mn_{6.5}Si_{10}B_{15}$ amorphous glass-coated wires having the total diameter of 14.5 μm and the metallic core diameter of 10.2 μm. The wire diameter is chosen such that it would be suitable for microwave applications. For this alloy composition, $\lambda \approx -2 \cdot 10^{-7}$.[23] For dc magnetization measurements, the wire length was 6 cm and the stress was applied by hanging a load of $1-10$ g at the wire end. The longitudinal magnetization curves under the effect of stress are shown in Fig. 2. The original curve for an unloaded wire shows a steep hysteresis but with a rather small remanence-to-saturation value $M_r/M_s = 0.2$, which corresponds to almost circumferential anisotropy with a small helical angle $\alpha \approx 12^0$ (deviation from the circumferential direction). Applying the tensile stress strengthens the circumferential anisotropy: the magnetization curve becomes almost linear with $M_r/M_s \approx 0$, the hysteresis disappears and the effective anisotropy field $H_K$ increases. For the unloaded wire $H_K = 2.6$ Oe, and it becomes 9 Oe, when a load of 8.5 g is applied. This magnetization behavior corresponds to the magnetostrictive energy given by equation (3) with parameters $\sigma_1 = 200$ MPa, $\sigma_2 = 44$ MPa provided that $\lambda = -2 \cdot 10^{-7}$ and load of 1g corresponds to a tension of 65 MPa. The applied stress was estimated considering different Young's modulus of the metallic core ($E_m$) and glass coating ($E_{gl}$) as $\sigma_{ex} = P k/(k S_m + S_{gl})$, where $P$ is the mechanical load, $k = E_m/E_{gl}$ and $S_m$ and $S_{gl}$ are areas of the metallic core and glass coating, respectively.

The complex-valued impedance $Z = V/i$ is found by measuring $S_{11}$-parameter (reflection coefficient) by means of a Hewlett-Packard 8753E Vector Network Analyser (VNA) with a specially designed microstripe cell (see Fig. 3) to minimize the post calibration mismatches originated by bonding the sample. Rather than attempting to quantify the contribution of this error in the wire impedance, our approach is to minimize its effect. The microstripe portion connected to VNA is pre-calibrated with a standard 50 Ω load, which is then removed (Load "1" in Fig. 3). During the measurements, the microstripe cell with the bonded sample is matched to 50 Ω load (Load "2"). The proposed calibration of the microstripe line with a length of 3.5 cm holding the bonded wire of 1.4 cm ensures reliable impedance measurements up to frequencies of about 3 GHz. The wire is loaded with a weight in its middle (in this case 1g load corresponds to a much higher tension estimated to be 320 MPa, see Caption for Fig. 3). The cell with the sample is placed into the Helmholtz coil producing a dc magnetic field.

Figure 4 shows plots of impedance vs. applied tensile stress for two frequencies 500 MHz and 2.5 GHz with $H_{ex}$ as a parameter. If no field is applied, the stress effect is small being about 12% at 500 MHz and almost not noticeable at GHz frequencies. In the presence of $H_{ex}$, the wire impedance shows large changes in response to the application of the tensile stress. The highest stress sensitivity of 100% and 60% per 180 MPa, respectively for frequency 500 MHz and 2.5 GHz, is obtained for $H_{ex} = 3$ Oe that is about the same value as the anisotropy field for the unloaded wire.

This impedance vs. stress behavior is in agreement with the previous discussion based on the analysis of Eqs. (2)-(4). Since a negative magnetostriction wire has a nearly circumferential anisotropy (if no large frozen-in torsion exists), the applied tensile stress alone will not cause the change in $\mathbf{M}_0$ direction, and as a result, will not produce noticeable changes in the impedance at high frequencies. Applying the field $H_{ex}$ of the order of the anisotropy field saturates the wire in the axial direction. The tensile stress which enlarges the circumferential anisotropy in the case of negative magnetostriction acts in opposite way and rotates the magnetization back to the circular direction. The highest stress sensitivity is obtained for $H_{ex} \approx H_K$ (the anisotropy field for the unloaded wire), which is sufficient to saturate the wire. Further increase in $H_{ex}$ is unnecessary since a larger stress will be required to return the magnetization back. Such impedance vs. stress behavior can be quantitatively modeled using Eqs. (3) and (4) and assuming that the dynamical permeability $\mu$ in Eq. (3) is related with a uniform precession of the magnetization around the equilibrium magnetization $\mathbf{M}_0$. The theoretical plots showing the normalized impedance $Z(\sigma_{ex})$ with $H_{ex}$ as a parameter for frequency of 2.5 GHz are shown in Fig. 5, demonstrating very good quantitative agreement with the experimental results.

**IV. Stress-dependent permittivity of composite materials with the ferromagnetic wire inclusions**

The effect of stress on microwave MI in amorphous fine wires is proposed to utilize for the creation of a new stress-tunable composite medium for remote stress monitoring by microwave spectroscopy technology. The composite material consists of short wire inclusions (exhibiting stress-impedance effect) embedded into a dielectric matrix with the permittivity $\varepsilon$. The wire inclusions behave as electric dipole "scatterers". This dipole response of the composite system can be characterized by some effective permittivity $\varepsilon_{eff}$. Then, $\varepsilon_{eff}$ can have a resonance or relaxation dispersion (depending on the wire impedance) seen near the antenna resonance for an individual wire. If the skin effect in wires is strong ($a \gg \delta$), the effective dielectric response is determined by the wire geometry and permittivity of host material. In a general case of a moderate skin effect ($a \approx \delta$), the dispersion characteristics of $\varepsilon_{eff}$ depend on the surface impedance. Therefore, in composite containing ferromagnetic wires exhibiting large MI effect at microwave frequencies, the effective permittivity may depend on an applied magnetic field[14,15] or internal stresses via the corresponding dependence of the longitudinal impedance.

The theoretical approach for such tunable composite materials was developed in our previous work.[15] The wire inclusions are polarized with an electric field of the incident electromagnetic wave $\mathbf{e}_0$. This field will excite the axial current which will have a nontrivial distribution along the wire when the wavelength $\lambda$ inside the material is of the order or smaller than the wire length. For the purpose to describe the external scattered field, the induced current can be regarded as a linear current in the antenna approximation: $a \ll \lambda$ and $a \ll l$. The current distribution inside the conductor is taken into account via the impedance boundary conditions of the form (1). Using the continuity equation $\partial i / \partial z = j\omega \, q(z)$, where $i(z)$ and $q(z)$ are linear densities of the induced current and charge, respectively, and imposing the zero boundary conditions for $i(z)$ at the wire ends ($i(-l/2) = i(l/2) \equiv 0$), the wire dipole moment $P_z$ is found by integrating the current along the wire:[15]

$$P_z = \frac{j}{\omega} \int_{-l/2}^{l/2} i(z) dz \tag{5}$$

The wire dielectric polarisability is then defined as $\chi_e = P_z /(e_z \pi \, a^2 l)$, where $\pi \, a^2 l$ is the volume occupied by the wire and $e_z$ is the projection of the local electrical field on the wire axis. At a very low inclusion volume concentration $p \ll p_c \propto 2a/l$, where $p_c$ is the percolation threshold for the wire-filled composites,[17] the effective permittivity $\varepsilon_{eff}$ can be represented as the dipole sum with the polarisability $<\chi_e>$ averaged over the inclusion orientations:

$$\varepsilon_{eff} = \varepsilon + 4\pi \, p < \chi_e > . \tag{6}$$

For such an approximation, the local electrical field **e** near a wire inclusion is assumed to be equal to the excitation field inside the composite matrix: $\mathbf{e} = \varepsilon \, \mathbf{e}_0$, where $\mathbf{e}_0$ is the electrical field of the electromagnetic wave outside of the composite sample. The interaction between wires can be taken into account within the mean field theory approach. Nevertheless, in the limit of small inclusion concentration any mean field theory reduces to Eq. (6) constituting "the ideal gas approximation".

In general, the current distribution $i(z)$ is defined by the integro-differential equation which accounts for all the losses including the radiation losses and internal losses (magnetic and resistive).[15] Its solution can be found by iteration process. The zero approximation represents well all the main features of the scattering by a magnetic wire when the internal losses (introduced by the impedance boundary conditions) are essential and the radioactive losses can be neglected:[15]

$$\frac{\partial^2 i(z)}{\partial z^2} + \tilde{k}^2 i(z) \approx \frac{j\omega \varepsilon\, e_{0z}}{2\ln(l/a)}, \quad i(\pm l/2) = 0, \tag{7}$$

$$\tilde{k} = k_0 \left(1 + \frac{jc\varsigma_{zz}}{\omega a \ln(l/a)}\right)^{1/2}, \quad k_0 = \frac{\omega \sqrt{\varepsilon}}{c}. \tag{8}$$

Here $k_0$ is the wave number in the dielectric matrix. As it follows from Eqs. (7),(8), the implementation of the impedance boundary condition leads to the renormalized wave number $\tilde{k}$ which now depends on $\varsigma_{zz}$. The solution of Eq. (7), satisfying to the boundary conditions $i(-l/2) = i(l/2) \equiv 0$, takes a simple form:

$$i(z) = \frac{j\omega \varepsilon\, e_{0z}}{2\ln(l/a)\tilde{k}^2} \frac{\left(\cos(\tilde{k}l/2) - \cos(\tilde{k}z)\right)}{\cos(\tilde{k}l/2)}. \tag{9}$$

The function $\cos(\tilde{k}l/2)$ in the denominator of Eq. (9) provides a resonant current distribution at $\mathrm{Re}(\tilde{k}_{res})l = \pi(2n-1)$, where $n$ are integers, with the resonance wavelengths:

$$\lambda_{res,n} = \frac{2l}{2n-1} \sqrt{\varepsilon}\, \mathrm{Re}\left(1 + \frac{jc\varsigma_{zz}}{\omega a \ln(l/a)}\right)^{1/2}. \tag{10}$$

Integrating $i(z)$ in Eq. (9) with respect to $z$ and using Eq. (5), the dipole moment $P_z$ is given by:

$$P_z = \frac{\varepsilon\, e_{0z} l}{2\ln(l/a)\tilde{k}^2}\left(\frac{2}{\tilde{k}l}\tan(\tilde{k}l/2) - 1\right). \tag{11}$$

The induced polarisability $\chi_e = P_z/(e_z \pi a^2 l)$, where $e_z = \varepsilon\, e_{0z}$, has a dispersion with a characteristic resonance frequency $f_{res} = c/\lambda_{res,1}$ (the first resonance only should be considered as having the maximal intensity):

$$\chi_e = \frac{1}{2\pi (\tilde{k}a)^2 \ln(l/a)}\left(\frac{2\tan(\tilde{k}l/2)}{\tilde{k}l} - 1\right). \tag{12}$$

Using Eqs. (6) and (12), the resultant expression for $\varepsilon_{eff}$ is obtained in the limit of small inclusions concentrations:

$$\varepsilon_{eff} = \varepsilon + \left\langle \frac{2p}{(\tilde{k}a)^2 \ln(l/a)} \left( \frac{2\tan(\tilde{k}l/2)}{\tilde{k}l} - 1 \right) \right\rangle, \tag{13}$$

where the averaging $\langle ... \rangle$ is conducted over the inclusion orientations.

The second term in (8) has a small effect on the value of the resonance frequency which is determined by the usual half wavelength condition ($f_{res} = c/2l\sqrt{\varepsilon}$). However, it can greatly change the losses in the system resulting in $\varepsilon_{eff}$ experiencing a transformation from a resonant spectrum to a relaxation one when the wire impedance is changed by excess stresses or strains in the host material. The condition of a moderate skin effect appears to be essential to obtain a high sensitivity of $\varepsilon_{eff}$ to the wire magnetic structure. Substituting (3) into (7) gives:

$$\tilde{k} = k_0 \left( 1 + \frac{(1+j)}{2\ln(l/a)} \frac{\delta_0}{a} \sqrt{\mu} \cos^2(\theta) \right)^{1/2}. \tag{14}$$

Eq. (14) shows that for a strong skin effect ($\delta_0 \sqrt{\mu} \ll a$), the normalized wave number $\tilde{k}$ differs little from the wave number $k_0$. In this case, the relaxation in the system is mainly determined by the radiation which could be found by taking the first iteration of the general integro-differential equation for the current distribution.[15] Therefore, the magnetic parameters of the wire have no effect on the permittivity. In the opposite limit ($\delta_0 \sqrt{\mu} \gg a$), the internal losses are strong resulting in broadening of the dispersion behavior of $\varepsilon_{eff}$, which becomes insensitive to impedance change. Therefore, an essential dependence of $\varepsilon_{eff}$ on the wire magnetic structure can be expected at moderate skin effect ($\delta_0 \sqrt{\mu} \propto a$). It imposes restrictions on the wire diameter. For amorphous CoSiB wires with the resistivity of 130 $\mu\Omega\cdot$cm and the ac permeability shown in Fig. 1, the skin depth is about 1.5 microns at 2 GHz. The wire length $l$ is determined by the resonance frequency and can be adjusted by using a dielectric host with higher permittivity $\varepsilon \gg 1$. Some polymer or commercial epoxy (Shipley photoepoxy with $\varepsilon \sim 3$) can be used as a dielectric host. A fine-dispersion filler (powder) containing particles with a large polarisability can be used for further increasing $\varepsilon$, for example, BaTiO$_3$ ceramic microparticles or fine-disperse metal powder.[17]

Figure 6 shows the frequency plots for the effective permittivity with $H_{ex}$ and $\sigma_{ex}$ as parameters, calculated using Eq.(6) for a planar composite. In this modeling $\varepsilon = 16$ is taken, for which the antenna resonance frequency is 3.75 GHz for the wire length of 1 cm. The wire magnetic parameters are taken the same as those in the experiment. The stress-dependence effect shows up in changing the character of the dispersion curves. When no excess stresses exist in the material ($\sigma_{ex} = 0$) and in the absence of $H_{ex}$, the dispersion curves are of a resonance type: at $f = f_{res}$ the imaginary part has a peak ($\varepsilon''_{eff} = 14$) and the real part demonstrates the so-called anomalous dispersion when it goes from a maximum ($\varepsilon'_{eff} = 26$) to a minimum ($\varepsilon'_{eff} = 12$) when frequency is increased past the resonance. Applying a magnetic field $H_{ex} \approx H_K = 3$ Oe while $\sigma_{ex} = 0$ increases the impedance, and, as a consequence, the internal losses in the inclusion, which results in the dispersion of a relaxation type. In the presence of $H_{ex}$ the resonance frequency also slightly shifts towards higher frequencies. Then, if the external stress is also applied, the dispersion curves become again of a resonance type since the impedance decreases as a consequence of the magnetization rotation back to the circumferential direction. For $H_{ex} = 3$ Oe, applying a stress of 650 MPa increases the relaxation peak almost twice.

These large changes in $\varepsilon_{eff}$ under the effect of stress can be detected by microwave scanning techniques, in particular, microwave loss spectroscopy based on contrast mapping. In this method, a large contrast between the permittivity values of stressed and unstressed regions is used searching for substantial stress gradients. It means that difficult calibration problems can be avoided. For practical applications it would be preferable to have a stress-tunable composite material operating without any use of a dc magnetic field. Then, a special anisotropy of a helical type will be needed to realize stress-sensitive microwave MI. This can be achieved by annealing under torsion.[24,25]. The other possibilities are related with creep induced anisotropy which is in transverse direction with respect to that originated by the initial tensile stress (tending to axial in negative magnetostrictive wires and to circumferential in positive magnetostrictive wires).[26]

## V. Conclusion

The present paper proposes a new composite material containing short amorphous magnetic wires, the effective permittivity of which can be tuned by applying a stress. This effect is based on the dependence of the wire high frequency impedance on stress. It has been demonstrated that in CoMnSiB amorphous wires the impedance changes as much as 100% and 60% per 180 MPa for frequencies 500 MHz and 2.5 GHz, respectively. Using a theoretical analysis and magnetization loop measurements, this behavior is proven to owe mainly to the directional change in the equilibrium magnetization caused by the applied stress or magnetic field. The analysis of the dielectric response from such composite materials shows that the dispersion behavior of the effective permittivity can experience a transformation from a relaxation spectrum to a resonant one due to an excess stress in the material.

Since the mechanical stress and its distribution into a construction are static in nature, an additional mediate physical process is required to visualize them. The proposed sensing media can be used for structural stress monitoring by microwave spectroscopy. The composite material can be made as a bulk medium or as a thin cover to image the mechanical stress distribution inside construction or on its surface. A possible design of monitoring system could use a network of stress-sensitive composite "blocks" embedded into the structure. The monitoring could be organized by means of radar scanning over the whole structure, when a microwave beam subsequently interrogates all the embedded sensing blocks over some frequency range which includes the resonance frequency. The reflection resonance spectrum from each block would be proportional to the acting stress. The total stress distribution in the structure can be restored using the microwave spectroscopy based on contrast mapping. This method is free from the problems related with the drift of the composite parameters due to the temperature or some other external factors. This principle is similar to the differential sensor circuit when the same off-set signal in two channels does not influence on the differential output signal.

**Acknowledgment:**

The authors would like to thank Dr Vladimir Larin, the Director of MFTI LTD (Kishinev, Republic of Moldova, www.microwires.com), for providing the amorphous ferromagnetic glass-coated wires.

# Figures:

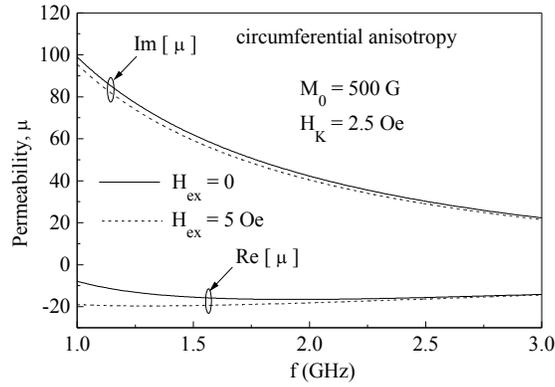

Fig.1. Dispersion curves of the effective permeability $\mu$ for $H_{ex}=0$ and $H_{ex}=5$ Oe in the GHz range (frequencies higher than the ferromagnetic resonance frequency for the parameters used $M_0 = 500$ G, $H_K = 2.5$ Oe).

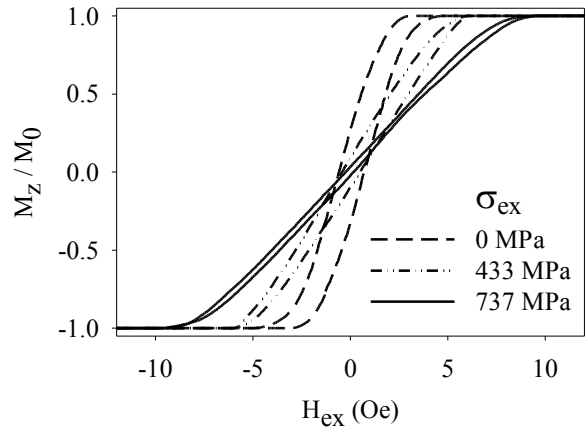

Fig. 2. Axial dc magnetization curves $M_z/M_0 = \cos(\theta)$ vs. $H_{ex}$ with the applied stress as a parameter. The wire length was 6 cm and the load was attached at the wire free end. In this case the tension created by 1g is estimated to equal 65 MPa.

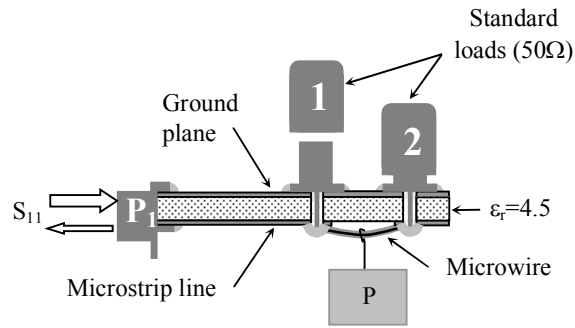

Fig. 3. Sketch of the microstripe cell with a wire sample for impedance measurement. The wire is loaded at the middle with a weight $P$ which imposes a stress of $\sigma_{ex} = P\,k/\bigl(2(kS_m + S_{gl})\sin(\psi)\bigr)$, where $\psi \approx 6^0$ is the angle between the bent wire and the horizontal direction.

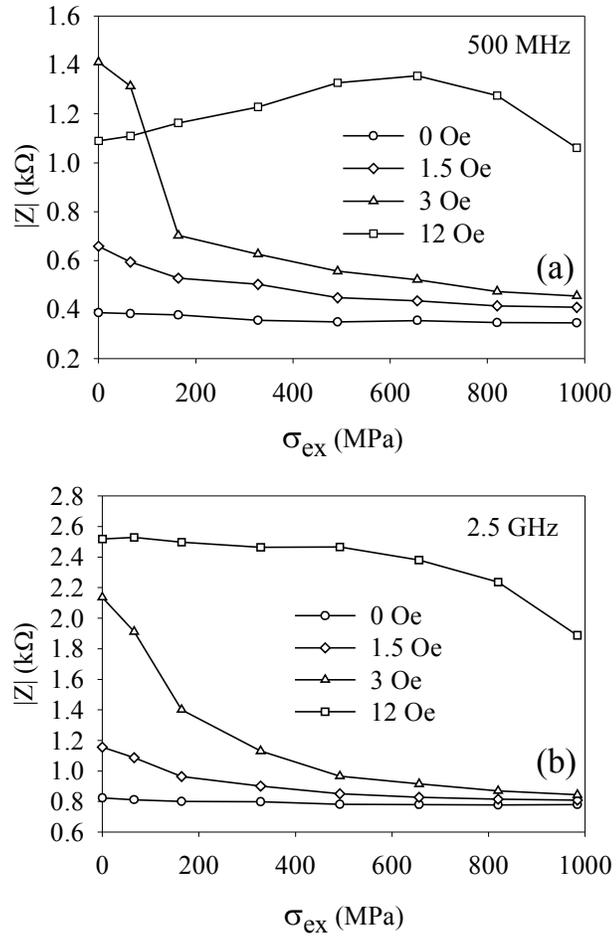

Fig. 4. Experimental plots of impedance vs. applied stress with the external magnetic field as a parameter for two frequencies: 500 MHz in (a) and 2.5 GHz in (b). The load of 1g applied at the middle of the wire is estimated to impose a stress of 320 MPa.

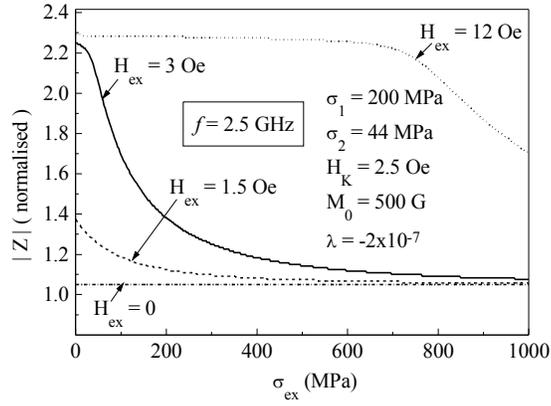

Fig. 5. Theoretical plots of impedance vs. applied stress with the external magnetic field as a parameter. The parameters used for calculation are $\lambda = -2 \cdot 10^{-7}$, $M_0 = 500$ G, $\sigma_1 = 200$ MPa, $\sigma_2 = 44$ MPa.

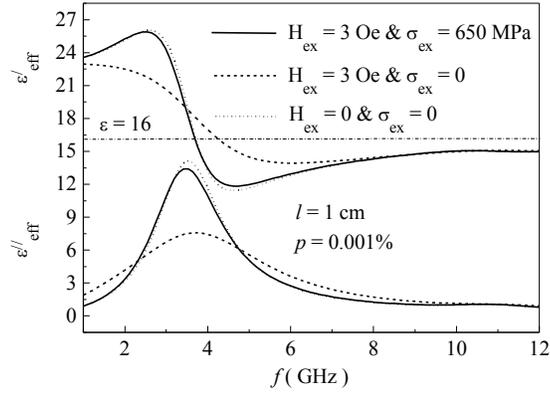

Fig. 6. Dispersion curves of the effective permittivity of composite material containing short MI wire inclusions with $H_{ex}$ and $\sigma_{ex}$ as parameters. The volume concentration of wires $p = 0.001\%$, the matrix permittivity $\varepsilon = 16$, and the wire length $l = 1$ cm. The magnetic parameters are same with those used for Fig. 5.